\documentclass[twocolumn,10pt]{IEEEtran}
\usepackage{cite}
\ifCLASSINFOpdf
\else
   \usepackage[dvips]{graphicx}
\fi

\usepackage{amsfonts}
\usepackage{lpic}
\usepackage{psfrag}
\usepackage{epsf}
\usepackage{dsfont}
\usepackage{textcomp}
\usepackage{stmaryrd}
\usepackage{amssymb}
\usepackage{mathrsfs}
\usepackage{textcomp}
\usepackage{hhline}
\usepackage{flushend}

\newcommand{\PNSNR}{\mathsf{SNR}^\mathrm{PN}}
\newcommand{\CHSNR}{\mathsf{SNR}^\mathrm{CH}}
\newcommand{\PN}{\mathrm{PN}}
\newcommand{\CH}{\mathrm{CH}}
\newcommand{\Ex}[2]{{\mathbb{E}_{#1}\left[#2\right]}} 	
\newcommand{\Es}{{E_\mathrm{s}}}

\usepackage[belowskip=-13pt,aboveskip=3pt,footnotesize]{caption}

\usepackage[cmex10]{amsmath}
\usepackage{accents}
\usepackage{array}
\usepackage{mdwmath}
\usepackage{mdwtab}
\begin{document}
%
\title{{\huge Oscillator Phase Noise and Small-Scale Channel Fading \\in Higher Frequency Bands}}
\author{\hspace{0in}\IEEEauthorblockN{M. Reza Khanzadi$^{\textrm{\dag,*}}$, Rajet Krishnan$^{\textrm{*}}$, Dan Kuylenstierna$^{\textrm{\dag}}$, and Thomas Eriksson$^{\textrm{*}}$}
\\	\IEEEauthorblockA{
\textdagger{Department of Microtechnology and Nanoscience, Microwave Electronics Laboratory}\\
\textasteriskcentered{Department of Signals and Systems, Communication Systems Group}
\\{Chalmers University of Technology, Gothenburg, Sweden}
\\\textit{\{khanzadi, rajet, dan.kuylenstierna, thomase\}@chalmers.se}} 
}

\maketitle

\begin{abstract}
This paper investigates the effect of oscillator phase noise and channel variations due to fading on the performance of communication systems at frequency bands higher than $10\mathrm{GHz}$. Phase noise and channel models are reviewed and technology-dependent bounds on the phase noise quality of radio oscillators are presented. Our study shows that, in general, both channel variations and phase noise can have severe effects on the system performance at high frequencies. Importantly, their relative severity depends on the application scenario and system parameters such as center frequency and bandwidth. Channel variations are seen to be more severe than phase noise when the relative velocity between the transmitter and receiver is high. On the other hand, performance degradation due to phase noise can be more severe when the center frequency is increased and the bandwidth is kept a constant, or when oscillators based on low power CMOS technology are used, as opposed to high power GaN HEMT based oscillators.
\end{abstract}
\IEEEpeerreviewmaketitle
\section{Introduction}

Scarcity of the microwave band motivates the need to move to higher frequency bands (greater than $10\mathrm{GHz}$) that enables access to several GHz of vacant spectrum \cite{Mehrpouyan2014_EBAND}. However, this transition to higher frequency bands presents new challenges, with channel variations and phase noise being identified as some of the most critical \cite{Mehrpouyan2014_EBAND}.

It is known that both channel variations due to mobility and phase noise in radio frequency oscillators increases with frequency \cite{clarke1968statistical},\cite{Mehrpouyan2014_EBAND}. Furthermore, both channel variations and the phase noise manifest as a multiplicative form of noise, in that, they multiply with the transmitted signal of interest~\cite{Krishnan2012_1}. Hence, when both channel variations and phase noise are present in a practical system, it is interesting to know which noise is more dominant in terms of its impact on the system performance. This knowledge is also useful for designing receiver algorithms, where a pertinent question is whether one needs to design separate or joint channel-phase noise compensation algorithms.

The goal of this paper is to study the effects of oscillator phase noise and small-scale channel variations due to mobility on the performance of communication systems when operating in higher frequency bands, e.g., above $10\mathrm{GHz}$. It is also of interest to see how these effects change with frequency. In particular it is investigated how oscillators in different technologies will be affected. First, we present a technology dependent lower bound that quantifies the quality of practical oscillators. This bound can be used to predict the phase noise process statistics in higher frequency bands. Then we analyze the effect of channel variations and phase noise on the signal-to-noise ratio (SNR) of a system. Specifically, we study two scenarios -- in the first scenario, the received signal is only affected by oscillator phase noise, and the channel is assumed to be known perfectly. In the second scenario, the received signal is considered to be affected only by the time-varying channel due to fading and phase noise is absent. For both scenarios, the received signal is assumed to be compensated by estimators that achieve the minimum mean square error. Then the impact of the residual error due to channel variations and phase noise on the SNR is analyzed separately. To this end, we derive the Modified Bayesian Cr\'amer Rao Bound (MBCRB) for the channel and phase noise estimators that are assumed to be used at the receiver.

Finally, we present extensive simulation results that analyze the effects of relative velocity, oscillator quality, operating center frequency and the bandwidth on the system performance. Based on our analysis, we conclude that channel variations due to fading and phase noise can have severe effects on the system performance at high frequencies, and their relative severity depends on the application scenario and system parameters like center frequency and bandwidth. Channel variations are seen to be more severe than phase noise when the relative velocity between the transmitter and receiver is high, and when the center frequency is increased along with the bandwidth of the system. On the other hand, performance degradation due to phase noise can be more severe when the center frequency is increased and the bandwidth is kept a constant. The severity of phase noise is also seen to depend heavily on the design technology of the oscillators -- when oscillators based on high power GaN HEMT based oscillators are used, phase noise is less of a problem compared to channel fading while for low power CMOS based oscillators phase noise may be an issue for high frequency communication systems.

%
%
%
%

{\let\thefootnote\relax\footnote{\emph{Notations:} Italic letters $(x)$ are scalar variables, boldface letters $(\mathbf{ x})$ are vectors, uppercase boldface letters $(\mathbf{X})$ are matrices, $([\mathbf{X}]_{a,b})$ denotes the $(a,b)^{th}$ entry of matrix $\mathbf{X}$, $\Ex{}{\cdot}$ denotes the statistical expectation operation, $\mathcal{N}(x;\mu,\sigma^2)$ and $\mathcal{CN}(x;\mu,\sigma^2)$ denote the real and complex Gaussian distribution with variable $x$, mean $\mu$, and variance $\sigma^2$, respectively; $\log(\cdot)$ denotes the natural logarithm, and $(\cdot)^*$ and $(\cdot)^T$ denote the conjugate and transpose, respectively.}}\par
\section{System Model}
Consider the transmission of a block of $K$ data symbols over a time-variant Rayleigh fading channel, affected by random oscillator phase noise. In the case of perfect timing and frequency synchronization, the received signal after sampling the output of the matched filter at Nyquist rate can be written as \cite{Krishnan2013_1} 
\begin{align}
\label{eq:System_Model_general}
y_k&= e^{\jmath\theta_k} h_k s_k +w_k,~k\in\{1,\dots,K\},
\end{align}
where $\theta_k$ represents the phase noise affecting the $k$th received signal due to noisy transmit and receive local oscillators. Furthermore, $h_k$ represents the complex channel coefficient at time instant $k$, and $w_k$ is a realization of a zero-mean complex circularly symmetric additive white Gaussian noise (AWGN) with variance~$\sigma^2_{\mathrm{w}}$. We denote the transmitted and received symbol sequences as~$\mathbf{y}=\{y_k\}_{k=1}^K$ and~$\mathbf{s}=\{s_k\}_{k=1}^K$, respectively.
%

%

In the sequel, we first present a detailed background on the Wiener phase noise model for $\theta_k$, and the Clarke's model for $ h_k$.
%
\subsection{Oscillator Phase Noise}
\begin{figure}[t]
\begin{center}
\psfrag{kap}[][][1.2]{{$\frac{\kappa}{f^2}$}}
\psfrag{Lf}[][][1]{{$\mathcal{L}(f)$}}
\psfrag{lev}[][][1.2]{{$\frac{\kappa}{(f_{3\mathrm{dB}})^2}$}}
\psfrag{logf}[][][0.8]{$\log_{10}(f)$}
\psfrag{f3dB}[][][0.8]{$f_{3\mathrm{dB}}$}
\includegraphics[width=1.8in]{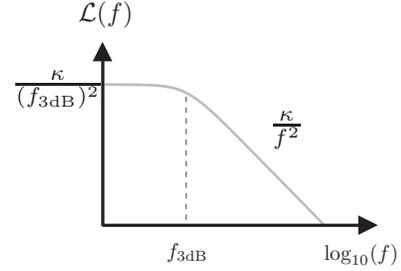}
\caption{The SSB spectrum of the oscillator in case of the Wiener phase noise. Here, $f$ denotes the offset frequency from the carrier, and $f_{3\mathrm{dB}}=\kappa \pi$.}
\label{fig:PN_OSC_PSD}
\end{center}
\end{figure}

Consider the case where the channel coefficient $h_k$ is perfectly known and compensated at the receiver. Assuming that $|h_k|=1$, the system model \eqref{eq:System_Model_general} can be rewritten as
\begin{align}
\label{eq:System_Model_PN}
y_k = e^{\jmath\theta_k} s_k +w_k,~k\in\{1,\dots,K\}.
\end{align}
The phase noise samples are modeled as a discrete Wiener process,
\begin{align}
\label{eq:Wiener_Model}
\theta_k=\theta_{k-1}+\zeta_{k-1},
\end{align}
where the phase noise innovation process $\zeta_{k}$ is a white zero-mean Gaussian random process, i.e.,~$\zeta_k\sim\mathcal{N}(0,\sigma^2_\zeta)$~\cite{Khanzadi2013_1}.\footnote{For discussions on the limitations of this model see \cite{Khanzadi2011,Khanzadi2013_2_ColoredPNEst} and references therein.} This discrete process corresponds to the sampled version of the continuous time Wiener process, which is the result of the sum of the phase noise processes at the transmit and receive oscillators. The samples are obtained at Nyquist rate in every $T_{\mathrm{s}}$ seconds, where $T_{\mathrm{s}}$ is the symbol interval.
Spectral measurements such as the single-side band (SSB) phase noise spectrum are the common figures for characterizing oscillators. The SSB phase noise spectrum is defined as the normalized power of the oscillator at offset frequencies from the carrier and it is reported in $\mathrm {dBc}/\mathrm {Hz}$. For Wiener phase noise, the SSB spectrum has a Lorentzian shape \cite{Khanzadi2011}
\begin{align}\label{eq:Lf_osc}
\mathcal{L}(f)=\frac{\kappa}{(\kappa \pi)^2+f^2},
\end{align}
where $f$ is the offset frequency (see Fig.~\ref{fig:PN_OSC_PSD}). This spectrum is fully characterized by a single parameter; the $3$dB single-sided bandwidth, $f_{3\mathrm{dB}}=\kappa \pi$ \cite[Sec.~V]{Chorti2006}, which corresponds to the frequency at which the noise power drops to half of the maximum noise level.
The connection between the continuous phase noise process and its discrete sampled version is captured by $\sigma^2_\zeta$, which is given as
\begin{align}\label{eq:innovation_var}
	\sigma^2_{\zeta} = \frac{4\pi f_{3\mathrm{dB}}}{BW} ,
\end{align}
where $BW=1/T_{\mathrm{s}}$ denotes the system bandwidth.
\subsection{Channel Fading}
In absence of oscillator phase noise, the input-output relation \eqref{eq:System_Model_general}  is  rewritten as
\begin{align}
\label{eq:System_Model_CH}
y_k = h_k s_k +w_k,~k\in\{1,\dots,K\}.
\end{align}
We consider a Rayleigh fading channel, which is an appropriate non-line-of-sight propagation model when there are many scattering objects in the environment. Based on Clarke's model \cite{clarke1968statistical}, the channel coefficients $h_k$ are modeled as zero-mean complex Gaussian random variables, i.e., $h_k\sim\mathcal{CN}(0,\sigma^2_{\mathrm{h}})$. Without loss of generality, we normalize the channel power by setting $\sigma^2_{\mathrm{h}}=1$. Upon splitting $h_k$ into its real $h^{(\mathrm{r})}_k$ and imaginary $h^{(\mathrm{i})}_k$ components, we obtain \cite{clarke1968statistical}
\begin{align}
&h_k=h^{(\mathrm{r})}_k+\jmath h^{(\mathrm{i})}_k\\
&h^{(\mathrm{r})}_k\sim\mathcal{N}(0,\frac{1}{2}),\quad h^{(\mathrm{i})}_k\sim\mathcal{N}(0,\frac{1}{2})\label{eq:Clark_Model_0}\\
&R_{h^{(\mathrm{r})}h^{(\mathrm{i})}}(\ell)=R_{h^{(\mathrm{i})}h^{(\mathrm{r})}}(\ell)=0\label{eq:Clark_Model_1}\\
&R_{h^{(\mathrm{r})}h^{(\mathrm{r})}}(\ell)=R_{h^{(\mathrm{i})}h^{(\mathrm{i})}}(\ell)=\frac{1}{2} \mathrm{J}_0(\frac{2\pi f_{\mathrm{D}}}{BW }|\ell|),\label{eq:Clark_Model_2}
\end{align}
where the function $ R_{x y}(\ell) = \Ex{}{x(k)y(k+\ell)}$ in \eqref{eq:Clark_Model_1} and \eqref{eq:Clark_Model_2} represents the correlation function between the random variables $x$ and $y$. In \eqref{eq:Clark_Model_2}, $\mathrm{J}_0$ is the zero-order Bessel function of the first kind, and $f_{\mathrm{D}}$ is the maximum Doppler frequency, given by
\begin{align}
	f_{\mathrm{D}}=\frac{v f_0}{c}
\end{align}
where $v$ is the relative speed between the transmitted and the receiver, $f_0$ is the center frequency of the radio frequency signal,  and $c=3\times 10^8~[\text{m}/\text{s}]$ is the speed of light. Note that the Doppler frequency scales linearly  with $f_0$.

In the next section we employ the models provided in \eqref{eq:System_Model_PN} and \eqref{eq:System_Model_CH} to evaluate the effect of phase noise and channel fading on the the performance of the system, where the performance metric considered is the SNR of the received signal.
\section{Effect of Parameter Estimation Errors on the SNR}
In this section, we investigate the effect of phase noise and channel fading on the SNR at the receiver.  As stated before, two scenarios are considered -- in the first scenario, an estimator is employed by the receiver to track the random time varying phase noise process. In the second scenario, an estimator is used to track the time varying channel fading process. For both the scenarios considered, the estimate of the parameter of interest is used to compensate its effect on the system performance. However, residual estimation errors remain, which influence the system performance. The SNR derived in this section corresponds to the SNR of the received signal after its compensation at the receiver.
%
\subsection{Oscillator Phase Noise}
Consider the system model in \eqref{eq:System_Model_PN}, and assume that the receiver employs a phase noise estimator that tracks the discrete phase noise process in \eqref{eq:Wiener_Model}. Specifically, this estimator tracks $\theta_k$ in each time instant, and let $\hat{\theta}_k$ be the estimate of $\theta_k$ in the $k$th time instant. In order to compensate the effect of phase noise on the received signal, it is rotated by~$-\hat{\theta}_k$ at the receiver,
\begin{align}
\label{eq:System_Model_PN_Derotate}
	 e^{-\jmath\hat{\theta}_k}y = e^{\jmath\epsilon_k} s_k +e^{-\jmath\hat{\theta}_k}w_k,
\end{align}
where~$\epsilon_k=\theta_k-\hat{\theta}_k$ denotes the estimation error, and~$e^{-\jmath\hat{\theta}_k}w_k$ has the same statistics as~$w_k$. We model~$\epsilon_k$ as a zero-mean Gaussian random variable, i.e.,~$\epsilon_k\sim\mathcal{N}(0,\sigma^2_{\epsilon,k})$ for $k=1,\dots,K$~\cite{Khanzadi2013_1}, where $\sigma^2_{\epsilon,k}$ indicates that the phase noise estimation variance depends on (position) index of the received signal in the block. Next we rewrite~\eqref{eq:System_Model_PN_Derotate} as
\begin{align}
\label{eq:System_Model_PN_Derotate_Re}
	e^{-\jmath\hat{\theta}_k}y = s_k + (e^{\jmath\epsilon_k} -1)s_k +e^{-\jmath\hat{\theta}_k}w_k,
\end{align}
where $(e^{\jmath\epsilon_k} -1)s_k$ represents additive noise term due to the residual phase noise estimation error. We now use \eqref{eq:System_Model_PN_Derotate_Re} to obtain the SNR at time instant $k$, which is written as the ratio of the desired signal power to the signal power due to AWGN and phase estimation error,
\begin{align}
\label{eq:PN_SNR}
\PNSNR_k&=\frac{\Ex{}{|s_k|^2}}{\Ex{}{2(1-\cos(\epsilon_k))|s_k|^2}+\Ex{}{|w'_k|^2}}\\
&=\frac{\Es}{2\Es(1-e^{-\frac{\sigma^2_{\epsilon,k}}{2}})+\sigma^2_{\mathrm{w}}}.
\end{align}
\subsection{Channel Fading}
We now analyze the effect of channel estimation errors on the SNR at the receiver. We consider a channel estimator at the receiver that provides an estimate of the instantaneous channel coefficient $h_k$, denoted as $\hat{h}_k$. The channel estimate is modeled as
\begin{align}
\label{eq:Channel_Est_Def}
\hat{h}_k=h_k+\varepsilon_k
\end{align}
where $\varepsilon_k\sim\mathcal{CN}(0,\sigma^2_{\varepsilon,k})$. In order to compensate for the effect of channel, we multiply the received signal \eqref{eq:System_Model_CH} by the conjugate of the channel estimate as
\begin{align}
\hat{h}_k^*y_k &= \hat{h}_k^*h_k s_k +\hat{h}_k^*w_k\\
					&= (h_k^*+\varepsilon_k^*) h_k s_k +(h_k^*+\varepsilon_k^*)w_k\label{eq:System_Model_CH_Compensate}\\
					&= |h_k|^2s_k+\varepsilon_k^* h_k s_k +(h_k^*+\varepsilon_k^*)w_k.
\end{align}
In \eqref{eq:System_Model_CH_Compensate}, we have substituted $\hat{h}_k$ from \eqref{eq:Channel_Est_Def}. Using \eqref{eq:System_Model_CH_Compensate} the SNR for $k$th symbol of the block is obtained as follows
\begin{align}
\hspace{-0.4cm}\CHSNR_k&=\frac{\Ex{}{|h_k|^4|s_k|^2}}{\Ex{}{|\varepsilon_k|^2 |h_k|^2 |s_k|^2}+\Ex{}{(|h_k|^2+|\varepsilon_k|^2)|w_k|^2}}\\
&=\frac{\Es}{\sigma^2_{\varepsilon,k}(\Es+\sigma^2_{\mathrm{w}})+\sigma^2_{\mathrm{w}}}.
\label{eq:CH_SNR}
\end{align}

As we observe from \eqref{eq:PN_SNR} and \eqref{eq:CH_SNR}, the SNR after estimation of phase noise and channel fading depends on the variance of estimation errors. In the next section we provide lower bounds on the estimation error variance for each scenario.

\section{Lower Bound on Estimation Error Variance}
In order to assess the estimation performance of a random parameter, the Bayesian Cram\'{e}r-Rao bound (BCRB) can be utilized -- this bound gives a tight lower bound on the mean square error (MSE) of the estimator of interest \cite{Andrea1994}. Consider a burst-transmission system, where $K$ symbols, denoted by the vector $\mathbf{s} = [s_{1}, \ldots,, s_{K}]^{\mathrm{T}}$, is transmitted in each burst.  According to the system model \eqref{eq:System_Model_PN}, a frame of signals $\mathbf{y}$ is received with the phase distorted by a vector of oscillator phase noise denoted by $\boldsymbol{\theta}= [\theta_1, \ldots, \theta_K]^{\mathrm{T}}$, with its prior probability density function (pdf) denoted by $f(\boldsymbol{\theta})$. The BCRB satisfies the following inequality for the MSE associated with a phase noise estimator:
\begin{align}
\label{BCRB_def}
&\mathbb{E}_{\mathbf{y},\boldsymbol{\theta}}\left[\left(\hat{\boldsymbol{\theta}}-\boldsymbol{\theta}\right)\left(\hat{\boldsymbol{\theta}}-\boldsymbol{\theta}\right)^T\right] - \mathbf{B}^{-1}_\PN \succeq \boldsymbol{0},\nonumber\\
&\mathbf{B}_\PN=\mathbb{E}_{\boldsymbol{\theta}}\left[\mathbf{F}(\boldsymbol{\theta})\right]+\mathbb{E}_{\boldsymbol{\theta}}\left[-\frac{\partial^2}{\partial\boldsymbol{\theta}^2} \log f(\boldsymbol{\theta})\right],
\end{align}
where $\hat{\boldsymbol{\theta}}$ denotes an estimator of $\boldsymbol{\theta}$, $\mathbf{B}_\PN$ is the Bayesian information matrix (BIM), and for a matrix $\mathbf{Z}$, $\mathbf{Z}\succeq \boldsymbol{0}$ implies that $\mathbf{Z}$  is positive semi-definite. In \eqref{BCRB_def}, $\mathbf{F}(\boldsymbol{\theta})$ is defined as
\begin{align}
\label{FIM_MBCRB}
&\mathbf{F}(\boldsymbol{\theta})=\mathbb{E}_{\mathbf{s}}\left[\mathbb{E}_{\mathbf{y}|\boldsymbol{\theta},\mathbf{s}}\left[-\frac{\partial^2}{\partial\boldsymbol{\theta}^2}  \log f(\mathbf{y}|\boldsymbol{\theta},\mathbf{s})\right]\right],
\end{align}
and this is referred to as the modified Fisher information matrix (FIM) \cite{Andrea1994}. Equivalently, the bound computed from (\ref{BCRB_def}) is called the modified Bayesian Cram\'{e}r-Rao bound (MBCRB). The MBCRB is a tight lower bound for non-data-aided parameter estimation at moderate and high SNR \cite{Khanzadi2013_1}. Note that in (\ref{BCRB_def}), the diagonal elements of $\mathbf{B}_\PN^{-1}$ provide a lower bound on the variance of the estimator for the elements in $\boldsymbol{\theta}$, i.e.,
\begin{align}
\label{error_def}
\sigma^2_{\epsilon,k}\triangleq&\mathbb{E}\Big[(\underbrace{{\theta_k}-\hat{{\theta}}_k}_{\triangleq\epsilon_k})^2\Big]\geq \left[\mathbf{B}_\PN^{-1}\right]_{k,k}.
\end{align}

From (\ref{BCRB_def})-(\ref{error_def}), we observe that the estimation error variance is entirely determined by $f(\boldsymbol{\theta})$ and $f(\mathbf{y}|\boldsymbol{\theta},\mathbf{s})$, which is the conditional pdf of the received signal $\mathbf{y}$ given $\boldsymbol{\theta}$ and $\mathbf{s}$  (usually referred to as the likelihood of $\boldsymbol{\theta}$).

For the phase noise model, where the phase noise innovations $\zeta_k$, for $k \in \{1,\ldots,K\}$, are correlated, $\mathbf{B}_\PN$ can be found in \cite[Eq.~22]{Khanzadi2013_1}. By adopting that result to the Wiener phase noise model in \eqref{eq:Wiener_Model}, where the phase noise innovations are uncorrelated, we obtain
\begin{align}
\label{PN_MBCRB_final}
\mathbf{B}_\PN=\frac{2E_s}{\sigma^2_{\mathrm{w}}} \mathbf{I}+\mathbf{C}^{-1},
\end{align}
where $\mathbf{I}$ is an $K\times K$ identity matrix, and
\begin{align}
\label{CovMatrix}
&[\mathbf{C}]_{m,n}=\sigma^2_{\theta_1}+(\min(m,n)-1)\sigma^2_\zeta\\
&m,n\in\{1\dots K\}\notag.
\end{align}
Here in \eqref{CovMatrix}, $\sigma^2_{\theta_1}$ denotes the phase noise variance associated with the first received signal in the block, where $\theta_1$ is uniformly distributed over $[0,2\pi)$.

Assuming that the phase noise estimator used at the receiver achieves an MSE performance close to the MBCRB and by substituting \eqref{PN_MBCRB_final} in \eqref{error_def}, then \eqref{error_def} in \eqref{eq:PN_SNR}, the SNR for the received signal model in \eqref{eq:System_Model_PN} after PN compensation is determined as
\begin{align}
\label{PN_SNR_Final}
\PNSNR_k&=\frac{\Es}{2\Es\left(1-\exp\left(-0.5[\mathbf{B}_\PN^{-1}]_{k,k}\right)\right)+\sigma^2_{\mathrm{w}}}.
\end{align}

Next we obtain the MBCRB for the channel estimator. First, we decompose the complex channel coefficient into its real and imaginary components and then calculate the MBCRB for the joint estimation of these components. We denote $\tilde{\mathbf{h}}^{\mathrm{T}}=[\mathbf{h}_{\mathrm{r}}^{\mathrm{T}} \mathbf{h}_{\mathrm{i}}^{\mathrm{T}}]$, where $\mathbf{h}_{\mathrm{r}}^{\mathrm{T}} = [h_1^{(\mathrm{r})}, \ldots, h_K^{(\mathrm{r})}]$ and~$\mathbf{h}_{\mathrm{i}}^{\mathrm{T}} = [h_1^{(\mathrm{i})}, \ldots, h_K^{(\mathrm{i})}]$. The BIM and the FIM are defined as
\begin{align}
&\mathbf{B}_\CH=\mathbb{E}_{\tilde{\mathbf{h}}}\left[\mathbf{F}(\tilde{\mathbf{h}})\right]+\mathbb{E}_{\tilde{\mathbf{h}}}\left[-\frac{\partial^2}{\partial\tilde{\mathbf{h}}^2} \log f(\tilde{\mathbf{h}})\right]\label{CH_BCRB_def}\\
&\mathbf{F}(\tilde{\mathbf{h}})=\mathbb{E}_{\mathbf{s}}\left[\mathbb{E}_{\mathbf{y}|\tilde{\mathbf{h}},\mathbf{s}}\left[-\frac{\partial^2}{\partial\tilde{\mathbf{h}}^2}  \log f(\mathbf{y}|\tilde{\mathbf{h}},\mathbf{s})\right]\right]\label{CH_F_def}.
\end{align}
Now, it remains to determine the likelihood function, $f(\mathbf{y}| \tilde{\mathbf{h}},\mathbf{s})$, and the a prior distribution of $\tilde{\mathbf{h}}$, denoted by $f(\tilde{\mathbf{h}})$.

Given that $w_k$, $k\in \{1,\ldots,K\}$, are i.i.d. random variables, and $y_k$ only depends on $h^{(\mathrm{r})}_k$, $h^{(\mathrm{i})}_k$ and $s_k$ according to \eqref{eq:System_Model_CH}, the likelihood function is written as
\begin{align}
\label{CH_likelihood}
f(\mathbf{y}| \tilde{\mathbf{h}},\mathbf{s})&=\prod_{k=1}^Kf(y_k| \tilde{\mathbf{h}},\mathbf{s})=\prod_{k=1}^Kf(y_k|h^{(\mathrm{r})}_k,h^{(\mathrm{i})}_k,s_k),
\end{align}
where
\begin{align}
&f(y_k|h^{(\mathrm{r})}_k,h^{(\mathrm{i})}_k,s_k)=\notag\\
&\hspace{1.8cm}\frac{1}{\sigma_{\mathrm{w}}^2\pi}\exp{\left(-\frac{|y_k-s_k (h_k^{(r)}+\jmath h_k^{(i)})|^2}{\sigma_{\mathrm{w}}^2}\right)}.
\end{align}
By substituting \eqref{CH_likelihood} in \eqref{CH_F_def}, it is straightforward to show that
\begin{align}
\mathbf{F}(\tilde{\mathbf{h}})=\frac{2\Es}{\sigma^2_{\mathrm{w}}} \mathbf{I}_{({2K\times 2K})}.
\end{align}

In order to find the prior distribution $f(\tilde{\mathbf{h}})$, we use that the real and imaginary components of the channel are i.i.d. Gaussian random variables. By using \eqref{eq:Clark_Model_0}-\eqref{eq:Clark_Model_2}, we obtain that $f(\tilde{\mathbf{h}})=\mathcal{N}(\tilde{\mathbf{h}};\mathbf{0},\boldsymbol{\Sigma})$ where
\begin{align}
\boldsymbol{\Sigma}=
\left[
\begin{array}{c|c}
\mathbf{R} & \mathbf{0} \\ \hline
\mathbf{0} &\mathbf {R}
\end{array}\right]_{({2K\times 2K})}
\end{align}
\begin{align}
 [\mathbf{R}]_{m,n}=\frac{1}{2} \mathrm{J}_0(\frac{2\pi f_{\mathrm{D}}}{BW} |m-n|),\quad m,n\in\{1\dots K\}.
\end{align}
%
By setting $f(\tilde{\mathbf{h}})$ and $\mathbf{F}(\tilde{\mathbf{h}})$ in \eqref{CH_BCRB_def}, followed by straightforward simplifications, we obtain
\begin{align}
\label{eq:CH_MBCRB_final}
\mathbf{B}_\CH=\frac{2\Es}{\sigma^2_{\mathrm{w}}}
\left[
\begin{array}{c|c}
\mathbf{I} & \mathbf{0} \\ \hline
\mathbf{0} &\mathbf {I}
\end{array}\right]+\boldsymbol{\Sigma}^{-1}.
\end{align}
The estimation error variance of $h_k$ can be found as the sum of the error variances associated with $\mathbf{h}_{\mathrm{r}}$ and $\mathbf{h}_{\mathrm{i}}$,
\begin{align}
\label{eq:CH_Est_Err_Var}
\sigma^2_{\varepsilon,k}\geq \left[\mathbf{B}_\CH^{-1}\right]_{k,k}+\left[\mathbf{B}_\CH^{-1}\right]_{k+K,k+K}=2\left[\mathbf{B}_\CH^{-1}\right]_{k,k},
\end{align}
where the equality in \eqref{eq:CH_Est_Err_Var} is because $\mathbf{B}$ in \eqref{eq:CH_MBCRB_final} is symmetric. Finally, by assuming that the channel estimator used at the receiver attains the MBCRB, and by substituting \eqref{eq:CH_Est_Err_Var} in \eqref{eq:CH_SNR}, we obtain
\begin{align}
\label{CH_SNR_Final}
\CHSNR_k&=\frac{\Es}{2\left[\mathbf{B}_\CH^{-1}\right]_{k,k}(\Es+\sigma^2_{\mathrm{w}})+\sigma^2_{\mathrm{w}}}.
\end{align}

\section{Results and Discussions}
\begin{table}[t]
\caption{Oscillator Design Parameters}
\label{tab:technologies}
\centering
\begin{tabular}{ |l|c|c|c|l| }
  	\hline
  Technology & $V_\mathrm{c/d}$ & $I_\mathrm{c/d}~[mA]$ & $Q_0$ & References\\
	\hhline{|=|=|=|=|=|}
  Si CMOS 	& $1$	& $5$		&$15$		&	\cite{sicard2010_32nm_Microwind35,burghartz2003design}\\
  	\hline
  SiGe HBT 	& $2$	& $30$	&$15$		& \cite{harame1995sige,burghartz2003design}\\
  	\hline
  InGaP HBT & $5$	& $25$	&$40$		& \cite{Byoung2011,bahl2001high}\\
  	\hline
  GaN HEMT 	& $20$ & $40$	&$40$		& \cite{wangyield2014,bahl2001high}\\
 	\hline
  GaAs HEMT & $4$	& $25$	&$40$		& \cite{menozzi2004off,Lin2014_6inch,bahl2001high}\\
  	\hline
\end{tabular}
\end{table}
We start by providing realistic lower bounds on  the innovation variance for the Wiener phase noise model. By using \eqref{eq:innovation_var} and $f_{3\mathrm{dB}}=\kappa \pi$, and employing the lower bounds on $\kappa$ given in \cite[Eq.~5]{Horberg2014_PN} and \cite[Eq.~28]{everard2012simplified}, we obtain
\begin{align}\label{eq:innov_var_lower_bound}
	\sigma^2_{\zeta} \geq \frac{\pi^2 \times 19.496 \times 10^{-21}}{I_\mathrm{d} V_\mathrm{d} Q_0^2}\frac{f_0^2}{BW}.
\end{align}
where $f_0$ is the operating center frequency of the oscillator, $Q_0$ is the unloaded quality factor of the resonator inside the oscillator, and $I_\mathrm{d}$ and $V_\mathrm{d}$ denote the operating collector/drain current and safe operating voltage of the transistor inside the oscillator, respectively.\footnote{Note that notations $I_c$ and $V_c$ for simplicity refer also to drain current and voltage.} The safe operating voltage is normally about $1/3$ of the device breakdown voltage $V_\mathrm{B}$. Typical values of $Q_0$, $I_\mathrm{c}$ and $V_\mathrm{B}$ depend on the design technology of the oscillators. Tab.~\ref{tab:technologies} provides these parameters for the various design technologies. As observed from \eqref{eq:innov_var_lower_bound}, the phase noise innovation variance grows quadratically with the operating center frequency $f_0$ and decreases linearly with $BW$.
\begin{figure}[t]
\begin{center}
\psfrag{BWCS}[][][0.8][21]{\color{blue}$BW=1\mathrm{MHz}$}
\psfrag{BWRS}[][][0.8][10]{\color{blue}$BW=f_0/1000$}
\psfrag{BWCG}[][][0.8][21]{\color{red}$BW=1\mathrm{MHz}$}
\psfrag{BWRG}[][][0.8][10]{\color{red}$BW=f_0/1000$}
\includegraphics[width=3in]{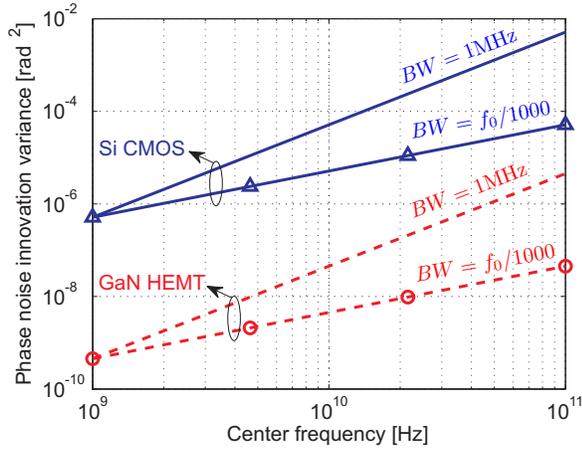}
\caption{Phase noise innovation variance $\sigma^2_\zeta$ for Si CMOS and GaN HEMT technologies versus the center frequency $f_0$ of the oscillator.}
\label{fig:WienerPN_Var_f1000BW}
\end{center}
\end{figure}
In Fig.~\ref{fig:WienerPN_Var_f1000BW} we compare the lower bound \eqref{eq:innov_var_lower_bound} for Si CMOS~\cite{okada201160} and GaN HEMT~\cite{lan2006GaN_HEMT_40GHz} technologies against different values of $f_0$. We consider two cases; in the first case a fixed bandwidth is used, $BW=1\mathrm{MHz}$. In the second case we linearly increase the bandwidth with $f_0$. Specifically, we set $BW=0.001 f_0$. We observe that for both the technologies and in the fixed bandwidth case, $\sigma^2_{\zeta}$ grows quadratically with $f_0$ ($20\mathrm{dB/dec}$). In the second case, $\sigma^2_{\zeta}$ scales almost linearly with $f_0$ ($10\mathrm{dB/dec}$). Furthermore, GaN HEMT technology has a lower $\sigma^2_{\zeta}$ than the Si CMOS technology for the scenarios considered. This difference is due to the higher quality factor obtained in GaN HEMT technology \cite{bahl2001high,burghartz2003design} and the higher available power \cite{wangyield2014}.

\begin{figure}[t]
\begin{center}
\psfrag{BWCS}[][][0.8][-42]{\color{blue}$BW=1\mathrm{MHz}$}
\psfrag{BWRS}[][][0.8][-6]{\color{blue}$BW=f_0/1000$}
\psfrag{BWCG}[][][0.8][0]{\color{red}$BW=1\mathrm{MHz}$}
\psfrag{BWRG}[][][0.8][0]{\color{red}$BW=f_0/1000$}
\includegraphics[width=3in]{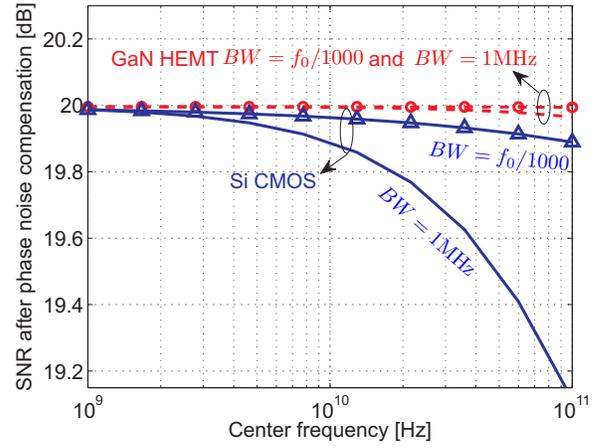}
\caption{The SNR after phase noise compensation for Si CMOS and GaN HEMT technologies versus the center frequency $f_0$ of the oscillator. Here, $\Es/\sigma^2_{\mathrm{w}}=20$dB, and $K=100$.}
\label{fig:WienerPN_SNR_Bounds}
\end{center}
\end{figure}

Fig.~\ref{fig:WienerPN_SNR_Bounds} illustrates the SNR after phase noise compensation for Si COM and GaN HEMT technologies. The SNR is calculated by using \eqref{PN_SNR_Final}, followed by an averaging operation over a block of $K=100$ symbols. For the Si CMOS technology, an SNR loss of $0.1\mathrm{dB}$ and $0.8\mathrm{dB}$ can be seen for $BW=1\mathrm{MHz}$ and $BW=0.001f_0$, respectively, when increasing  $f_0$ from $1\mathrm{GHz}$ to $100\mathrm{GHz}$. However, the SNR is less affected for the GaN HEMT technology.

\begin{figure}[t]
\begin{center}
\psfrag{BWCv1}[][][0.8][0]{\color{red}$BW=1\mathrm{MHz}$}
\psfrag{BWRv1}[][][0.8][0]{\color{red}$BW=f_0/1000$}
\psfrag{BWCv2}[][][0.8][0]{\color{blue}$BW=1\mathrm{MHz}$}
\psfrag{BWRv2}[][][0.8][0]{\color{blue}$BW=f_0/1000$}
\psfrag{v1}[][][0.8][0]{\color{blue}$v=50\mathrm{Km/h}$}
\psfrag{v2}[][][0.8][0]{\color{red}$v=1\mathrm{Km/h}$}
\includegraphics[width=3in]{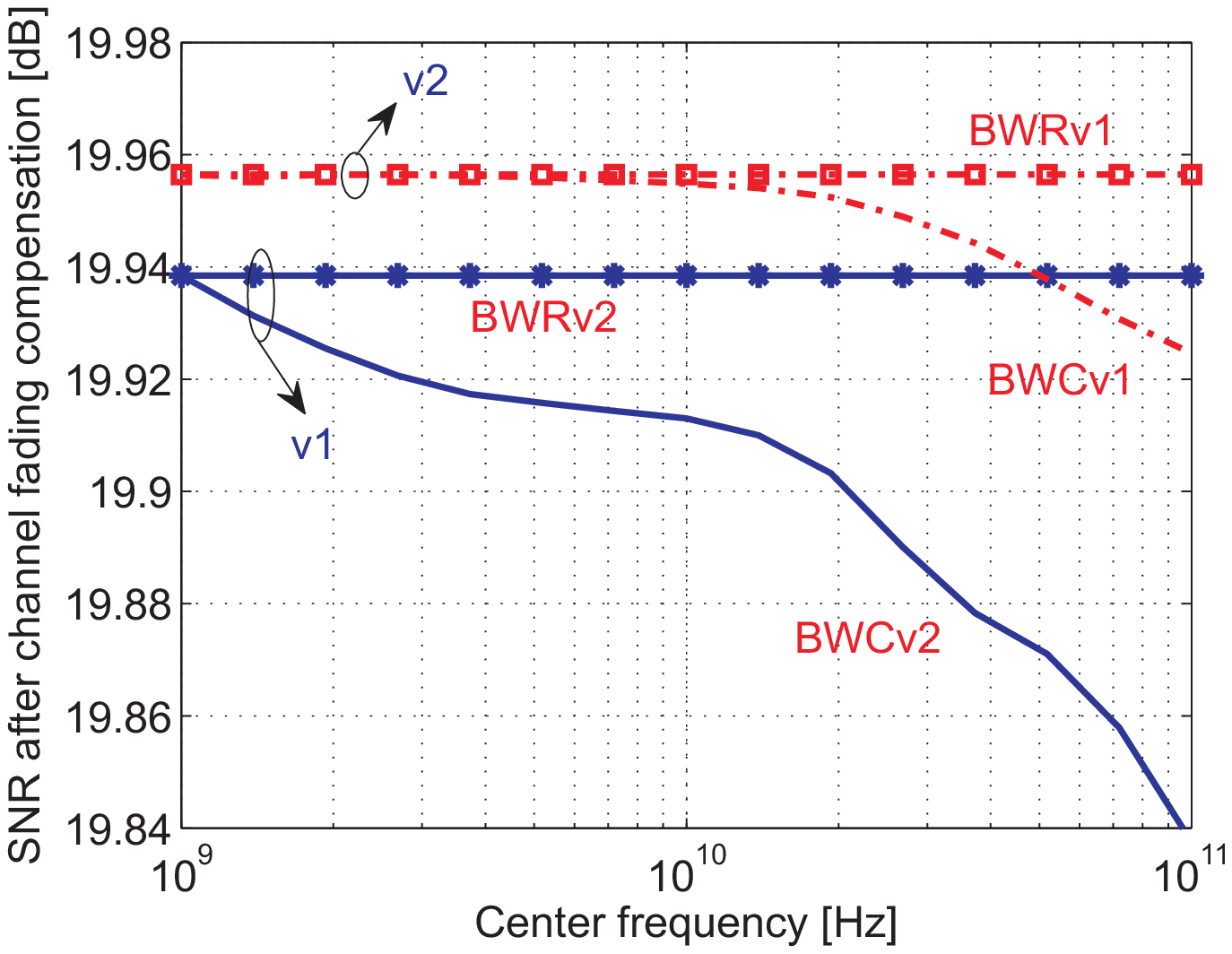}
\caption{The SNR after channel fading compensation for $1\mathrm{Km/h}$ and  $50\mathrm{Km/h}$ velocities versus the center frequency $f_0$. Here, $\Es/\sigma^2_{\mathrm{w}}=20$dB, and $K=100$.}
\label{fig:ChannelFading_SNR_Bounds}
\end{center}
\end{figure}

Fig.~\ref{fig:ChannelFading_SNR_Bounds} shows the SNR after channel fading compensation for relative velocities of $v=1\mathrm{Km/h}$ and $v=50\mathrm{Km/h}$. The SNR is calculated by using \eqref{CH_SNR_Final}, followed by an averaging operation over a block of $K=100$ symbols.\footnote{Note that $\boldsymbol{\Sigma}$ in \eqref{eq:CH_MBCRB_final} can be very close to a singular matrix that raises matrix inversion problems. To avoid this a constant bias value as explained in \cite{baddour2005autoregressive} is added to lag-zero of the channel's autocorrelation function.} When $BW=0.001f_0$, SNR stays constant for both relative velocities. This is because the autocorrelation function of the channel \eqref{eq:Clark_Model_2} stays constant.  On the other hand, when the  increasing $f_0$ with $BW=1\mathrm{MHz}$, SNR drops  $0.04\mathrm{dB}$ and $0.1\mathrm{dB}$ for $v=1\mathrm{Km/h}$ and $v=50\mathrm{Km/h}$, respectively. From figs.~\ref{fig:WienerPN_SNR_Bounds} and~\ref{fig:ChannelFading_SNR_Bounds}, we can clearly see that the degradation of the SNR due to phase noise is more severe than that due to the channel, when BW is a constant and an estimator that achieves MCRB is used at the receiver. This is because the phase noise innovation variance increases quadratically with $f_0$. However, the degradation of the SNR due to phase noise and the channel are seen to be similar when $BW$ scales with $f_0$.

\begin{figure}[t]
\begin{center}
\psfrag{L1}[][][0.8]{$f_{D}/BW=f_{3\mathrm{dB}}/BW$}
\psfrag{SNRPN}[][][0.8]{\color{blue} $\PNSNR$}
\psfrag{SNRCH}[][][0.8]{\color{red}$\CHSNR$}
\includegraphics[width=3in]{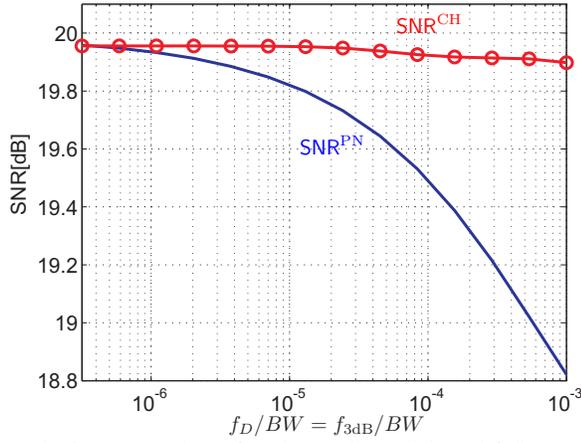}
\caption{The SNR comparison after phase noise and channel fading compensation when $f_{\mathrm{D}} /BW=f_{3\text{dB}}/BW$. Here, $\Es/\sigma^2_{\mathrm{w}}=20$dB, and $K=100$.}
\label{fig:SNR_vs_fDTs_f3dBTs}
\end{center}
\end{figure}

The channel fading based on the Clarke's model is a bandlimited process with single-side bandwidth given by $f_\mathrm{D}$. On the other hand, phase noise is not a bandlimited process -- it has infinite bandwidth. However, as mentioned before, we can define a $3\mathrm{dB}$ bandwidth for the phase noise process. In Fig.~\ref{fig:SNR_vs_fDTs_f3dBTs} we compare the effect of phase noise and channel fading on the SNR when $f_{3\mathrm{dB}}=f_{\mathrm{D}}$. It can be seen that in this particular comparison, phase noise affects the SNR more severely. We can also observe that the gap between the SNRs achieved in the scenarios considered dramatically grows upon increasing $f_{3\mathrm{dB}}$ and $f_{\mathrm{D}}$, while maintaining $f_{3\mathrm{dB}}=f_{\mathrm{D}}$. However, here it is worth noting that the $f_{3\mathrm{dB}}$ of most practical oscillators is significantly smaller than $f_{\mathrm{D}}$.

\begin{figure}[!h]
\begin{center}
\psfrag{eps}[][][0.8]{$\sigma^2_{\epsilon,k}=\sigma^2_{\varepsilon,k}$}
\psfrag{PNSNR}[][][0.8]{\color{blue} $\PNSNR_k$}
\psfrag{CHSNR}[][][0.8]{\color{red}$\CHSNR_k$}
\includegraphics[width=2.9in]{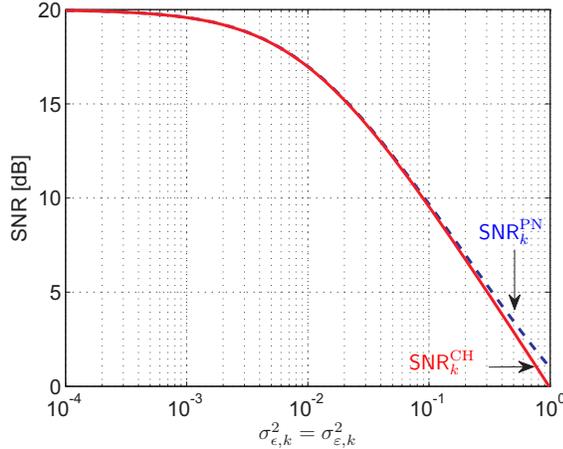}
\caption{The SNR comparison for equal error variance after compensation of phase noise and channel fading. Here, $\Es/\sigma^2_{\mathrm{w}}=20$dB.}
\label{fig:PNvsCH_GivenError}
\end{center}
\end{figure}

\begin{table}[b]
\caption{}
\label{tab:Standards}
\centering
\begin{tabular}{ |l|c|c|c|c| }
  	\hline
  Standard & $f_0[\mathrm{GHz}]$ & $BW[\mathrm{GHz}]$ & $\CHSNR [\mathrm{dB}]$ & $\PNSNR[\mathrm{dB}]$\\
	\hhline{|=|=|=|=|=|}
  IEEE 802.15.3c \cite{okada201160}	& $60$	& $2.16$		&$19.956$		&	$19.951$\\
  	\hline
  IEEE 802.11b \cite{zargari2004single}	& $2.4$	& $0.02$	   &$19.956$		& $19.952$\\
  	\hline
\end{tabular}
\end{table}
In Tab.~\ref{tab:Standards} we compare $\CHSNR$ and $\PNSNR$ for the IEEE~802.15.3c and IEEE~802.11b standards. In \cite{okada201160}, for the IEEE~802.15.3c standard a radio frequency oscillator with CMOS technology is used with $\mathcal{L}(1\mathrm{MHz})=-95\mathrm{dBc/Hz}$ in \eqref{eq:Lf_osc}. For the IEEE~802.11b standard, another CMOS-based oscillator with $\mathcal{L}(1\mathrm{MHz})=-115\mathrm{dBc/Hz}$ is employed in \cite{zargari2004single}. For a relative velocity of $v=0.5\mathrm{Km/h}$, the effects of channel fading and phase noise are observed to be of the same level, indicated by the identical SNRs achieved. This comparison shows that upon using better oscillators or when the relative velocity is slightly higher, channel fading has a more prominent effect on the performance compared to oscillator phase noise. Although the oscillator used in IEEE~802.11b has a lower phase noise level, we observe that $\PNSNR$ achieved for both the standards are similar. This is because $f_0/BW$ in IEEE~802.11b is $4.32$ times higher than that of IEEE~802.15.3c.

In Fig.~\ref{fig:PNvsCH_GivenError} we use \eqref{eq:PN_SNR} and \eqref{eq:CH_SNR} to compare the SNR degradation due to channel and phase noise estimation errors when $\sigma^2_{\epsilon,k}=\sigma^2_{\varepsilon,k}$. We observe that when the variance of the estimator increases, the SNR degradation due to channel fading is more severe for arbitrary estimation error variance.

\bibliographystyle{IEEEtran}
\bibliography{IEEEabrv,references}
\end{document}